\documentclass[journal=nalefd,manuscript=letter,layout=twocolumn]{achemso}

\usepackage{chemformula} 
\usepackage[T1]{fontenc} 
\usepackage{amssymb}
\author{Javier Osca}
\affiliation{IMEC, Kapeldreef 75, B-3001 Leuven, Belgium} 
\email{javier.osca@imec.be}
\author{Lloren\c{c} Serra}
\email{llorens.serra@uib.es}
\affiliation{IFISC (CSIC-UIB), E-07122 Palma, Spain} 
\alsoaffiliation{Department of Physics, University of the Balearic Islands, E-07122 Palma, Spain}
\title[Orbital]
  {Magnetic orbital motion and 0.5$e^2/h$ conductance of quantum-anomalous-Hall hybrid strips}

\abbreviations{QAH,SO}

\begin{document}


\begin{abstract}
The magnetic-induced  orbital motion of quasiparticles affects the conductance properties of 
a hybrid strip of a quantum-anomalous-Hall topological material 
with 
induced superconductivity. 
We elucidate the scenario of topological 
NSN ideal junctions in presence of orbital magnetic motion, showing how it 
leads to a halved quantized conductance $0.5e^2/h$
even in absence of Majorana modes. 
The magnetic orbital effect favours 
Fermionic
charged modes with finite wave numbers, in 
contradistinction to
Majorana zero modes which are chargeless zero-energy modes with
vanishing wave number.
The bias sensitivity of the 0.5 conductance plateau allows to distinguish 
the two cases. Conductance oscillations due to 
backscattering interference are absent in the charged Fermion case. 
\end{abstract}

\section{Introduction}

Quantum transport with devices of robust tunable properties is a very active 
field in Condensed Matter physics. The reasons are twofold: 
on the one hand, fundamental understanding of the 
often bizarre behaviors of a quantum system; on the other hand, 
practical  applications allowing to overcome present technological bottlenecks.
Specifically, we refer in this Letter to topological-transport systems \cite{Hass,Qi,Chiu},
able to  
host current-carrying states along the edges or boundaries of specific 
arrangements of different materials. The so-called Majorana zero modes in strips of
quantum-anomalous-Hall insulator (QAHI) material, with 
hybrid superconductivity,\cite{He294,Qi10,Chung,Wang13,Wang14,Wang15,Lian16,Lian18,Lian18b}
 clearly manifest the above dual interest:
they are electron-hole quantum superpositions that may allow
quantum computation through braiding.\cite{Nayak,Lian18}

Majorana zero modes have also been
studied in hybrid semiconductor-superconductor nanowires, where they remain
attached to the nanowire ends and yield  electrical zero-bias (anomalous) conductance 
peaks.\cite{Aguado,Lutrev}
 In contrast, Majorana modes in QAHI strips 
propagate along the lateral edges and they 
are characterized by a 
$0.5e^2/h$ conductance plateau when a single Majorana mode is active.
The corresponding  energy-momentum dependence
$\varepsilon(k)$ is linear, with the shape of a diagonal cross
centered around $k=0$ and $\varepsilon=0$.
These modes are chiral, with right- and left-movers localizing on opposite 
edges of the strip.
Interferometry with Majorana beam splitters was considered in Refs.\ \citenum{Akh09,Fu09,Roi18} 
while
quasiparticle reversal (crossed Andreev reflection) was discussed in Ref.\ \citenum{Rei13}.

In QAHI's the material magnetization is fundamental for the characterization of the 
topological phases; each phase having a topological invariant given by 
the number of zero-energy edge modes present. The internal magnetization is controlled by an 
external magnetic field acting like a trigger for the magnetization reversal; typically, 
sweeping $B_z$ from negative to positive 
all the system phases between the maximum (saturation)  magnetizations
in those opposite perpendicular orientations are explored. The external field
$B_z$ does not need to be strong, usually in 
the millitesla range, which is beneficial to retain superconductivity of the hybrid system.

Besides being a trigger for the internal magnetization, a perpendicular magnetic field also has  
a direct action on the charged electron and hole quasiparticles. 
The direct coupling of the field with the quasiparticles' spin can be neglected or absorbed in the similar coupling 
given by the internal magnetization. In addition, however, there is a direct 
influence of the external field on the orbital quasiparticle motion. This is the focus
of our present study. We will show how the orbital effect modifies the scenario of the
Majorana modes, leading to charged Fermionic modes 
with nonvanishing wavenumbers (as opposed to chargeless Majorana
zero modes with $k\approx 0$).

 For a 2D strip, infinite along $x$ and with lateral extension $L_y$ along $y$ 
 (see sketch in Fig.\ \ref{Fig1}a) the minimal substitution
$\vec{p}\to\vec{p}\pm \hbar y/\ell_z^2$, where $\ell_z^{-2}=eB_z/\hbar c$ is the 
magnetic length, describes the orbital field effect.
The $\pm$ signs yield the well known different motion of opposite charges in a given $B_z$ field. 
In a topological QAHI the $p^2$ term is rather small but, nevertheless, the above minimal substitution cannot
be neglected in the strong $p$-linear contribution of Rashba spin-orbit type.
Notice also that in a field of, say, $B_z\gtrsim  1$ mT it is 
$\ell_z\lesssim 1$ $\mu$m and, thus, it makes orbital effects again relevant
for strips of $L_y\gtrsim 1$ $\mu$m and fields in the millitesla range.

Remarkably, although the orbital effect alters the character
of the 
zero modes, 
the conductance still remains $0.5e^2/h$ in cases when a single 
charged mode propagates through the device.
We refer to a configuration such that 
the bias drop is symmetrical in the two terminals ($\pm V/2$) 
and the central superconductor is effectively grounded ($\mu=0$), with no current flowing
from it to the leads.
We conclude that a halved conductance plateau is not a unique signature of a Majorana mode, but it is also possible with Fermionic modes having
different  electron and hole transmissions such that one mode is transmitted and the other is reflected.

An essential difference between the above two scenarios is that the transport by a single charged Fermion requires a nonvanishing energy. As shown below, this causes the small-bias limit of the 0.5 conductance plateau to 
disappear in the charged Fermion case while it survives in the Majorana one.
Besides, for reduced transverse widths $L_y$, the Majorana transmission is
characterized by conductance 
oscillations.\cite{Osca18b} These oscillations are absent if transmission 
occurs through a charged Fermion mode in a perfect way.
It is worth noting here that 0.5 conductance plateaus in absence of Majorana modes
have also been suggested in Refs.\ \citenum{Huan18,Ji18} as a result of percolation through 
disordered QAHI strips. In this sense, we assume that disorder is 
small or moderate, such that the chiral edge character of the modes is still well defined throughout the device.
 
\section{Model and transport}

\begin{figure}[t]
\begin{center}
\includegraphics[width=0.45\textwidth,trim=1.2cm 4.7cm 2.7cm 3.6cm,clip]{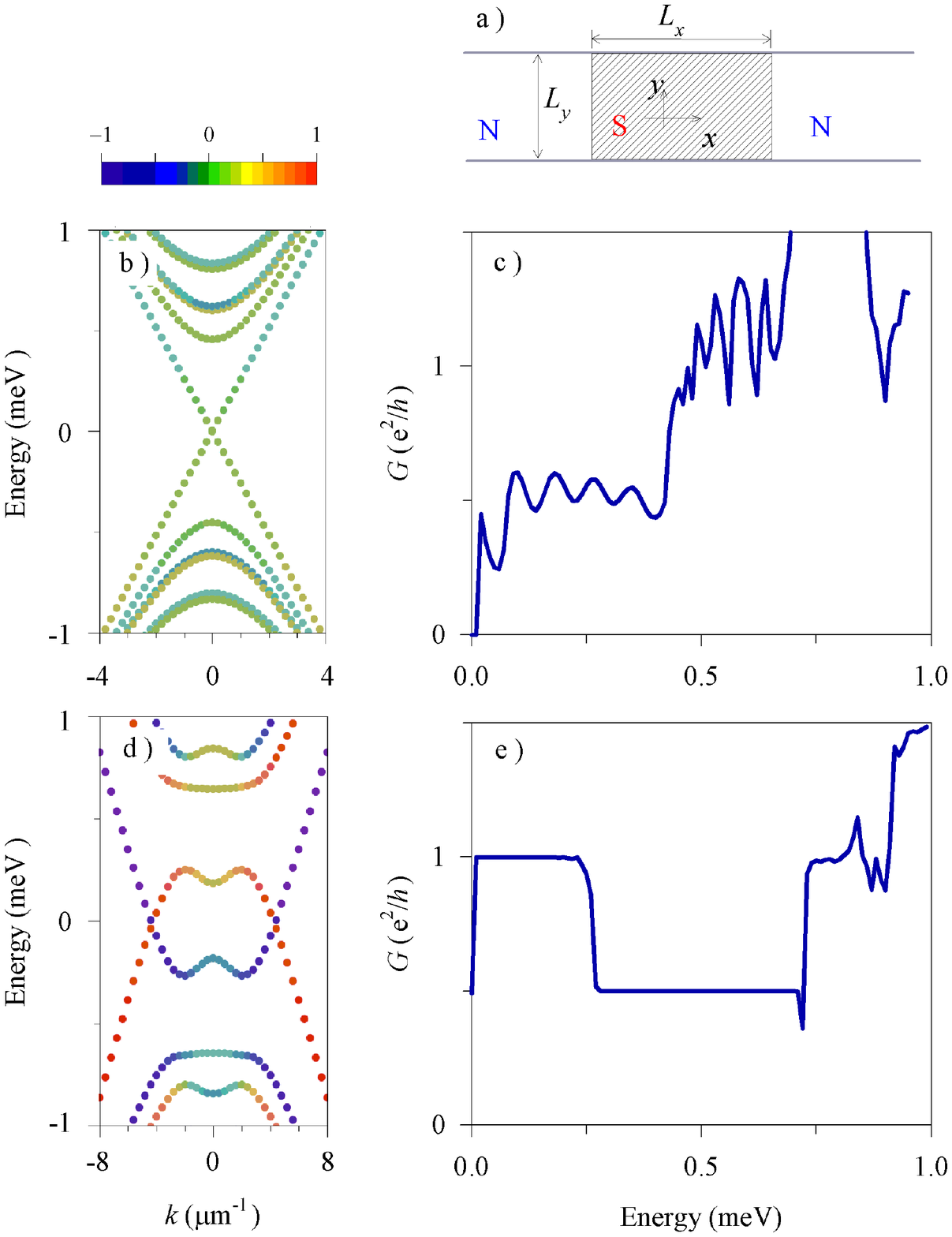}
\vspace{-3mm}
\end{center}
\caption{
Panels b), d): band structure of a hybrid strip of $L_y=3$ $\mu$m with b) a Majorana zero mode at $k=0$ 
and d) two charged Fermionic modes at finite $k$'s.
The charge of each state is given by the symbol color scale.
The Zeeman and magnetic length
parameters are $\Delta_Z=1.2$ meV and $\ell_z^{-2}=0.1$ $\mu{\rm m}^{-2}$ for b); and
 $\Delta_Z=1.5$ meV and $\ell_z^{-2}=4$ $\mu{\rm m}^{-2}$ for d).
 The right panels c) and e) show the bias-dependent conductances of an 
 NSN double junction with  $L_x=10$ $\mu$m (a) such that the central region is described by  the band structure shown on its corresponding left panel.
Other parameters: $m_0=1$ meV, $m_1=10^{-3}\, {\rm meV}\mu{\rm m}^2/\hbar^2$, 
$\alpha=0.26\, {\rm meV}\mu{\rm m}$,
$\Delta_{1,2}=(1,0.1)$ meV.
}
\vspace{-5mm}
\label{Fig1}
\end{figure}

We use the model of double-layer 2D hybrid QAHI strip  
of Refs.\ \citenum{He294,Qi10,Chung,Wang13,Wang14,Wang15,Lian16}.
More specifically, with the same notation and parameter 
definitions of Ref.\ \citenum{Osca18b}
the Hamiltonian reads
\begin{eqnarray}
{\cal H} &=& 
\left[\, m_0 + m_1 \left(p_x^2 +p_y^2\right)\, \right] \tau_z\, \lambda_x 
+ \Delta_Z\, \sigma_z \nonumber\\
&-& \frac{\alpha}{\hbar}\, \left(\,p_x\sigma_y-p_y\sigma_x\right)\, \tau_z\,\lambda_z \nonumber\\
&+& \Delta_p\, \tau_x + 
\Delta_m\, \tau_x\,\lambda_z\; .
\label{eq1}
\end{eqnarray}
Usual spin, electron-hole isospin and layer pseudospin are represented by 
$\sigma$, $\tau$ and $\lambda$ Pauli matrices, respectively.
The superconductor energy-gap parameter in the two layers $\Delta_{1,2}$ 
define the above $\Delta_{p,m}$ parameters by $\Delta_{p,m}=(\Delta_1\pm\Delta_2)/2$.
In an NSN double junction these parameters are nonzero only in the 
central S section of length $L_x$, as sketched in Fig.\ \ref{Fig1}a.  
$\Delta_Z$ represents the Zeeman-like term due to the magnetization 
of the material and $\alpha$ the spin-orbit coupling.
Typical values of the Hamiltonian 
parameters are discussed in Ref.\ \citenum{Osca18b}.
In particular, the 
energy and length scales are meV and $\mu$m, respectively.

With the minimal substitution, 
three new orbital terms are  added to 
Eq.\ (\ref{eq1}),
\begin{eqnarray}
{\cal H}_{\it orb} &=&
m_1\left( 
\frac{\hbar^2}{\ell_z^4}\, y^2\, \tau_z
-2 \frac{\hbar^2}{\ell_z^2}\, y\, p_x\right)\lambda_x\nonumber\\
&+& \frac{\alpha}{\ell_z^2}\, y\,\sigma_y \lambda_z\; .
\label{eq2}
\end{eqnarray} 
The first two contributions in Eq.\ (\ref{eq2}) originate in the $m_1$ 
quadratic-in-momentum term and they are familiar from the
kinetic contributions of the standard quantum Hall effect; 
the last one originates in the spin-orbit-like $\alpha$ term. 
We shall keep all three terms of Eq.\ (\ref{eq2}), although, 
in the present context of QAHI parameters 
the smallness of $m_1$ leads to a dominant magnetic $\alpha$-term
modification.
Note also that with our  energy (meV) and length ($\mu$m) units
the conversion between $\ell_z^{-2}$ and $B_z$ in milliteslas is
$B_z/{\rm mT} \approx 0.7\, \ell_z^{-2}/\mu{\rm m}^{-2}$; that is,
$\ell_z^{-2}=1\, \mu{\rm m}^{-2}$ corresponds to $B_z=0.7\, {\rm mT}$.

The formalism of phase-coherent transport in hybrid superconducting nanostructures was
reviewed in Ref.\ \citenum{Lamb98}. In particular, for our present purposes, the 
current-voltage relations in the two terminals $i=1,2$ read
\begin{equation}
\left(
\begin{array}{c}
I_1\\
I_2
\end{array}
\right)
=
\left(
\begin{array}{cc}
a_{11} & a_{12}\\
a_{21} & a_{22}
\end{array}
\right)
\left(
\begin{array}{c}
V_1\\
V_2
\end{array}
\right)\; ,
\label{eq3n}
\end{equation}
where the conductance matrix is given in terms of the number of propagating modes
in each terminal 
$N_i^\alpha(E)$ and the transmission probabilities between terminals $P_{ij}^{\alpha\beta}(E)$ (quasiparticle e/h type 
is indicated by $\alpha,\beta=\pm$) as
\begin{equation}
a_{ij}= \delta_{ij}N_i^+(E_i)+P_{ij}^{-+}(E_j)-P_{ij}^{++}(E_j)\; .
\label{eq4}
\end{equation} 
The energies $E_i$ in Eq.\ (\ref{eq4}) are related to the terminal-superconductor biases $E_i=eV_i-\mu$.
In a setup with a grounded superconductor and a symmetrical bias drop $V_{1,2}=\pm V/2$
the conductance is given by\cite{Lim12}
\begin{eqnarray}
G(E) =
\frac{e^2}{2h}
\left[\rule{0cm}{0.5cm}\right.
P_{11}^{-+}(E)
&+&P_{11}^{+-}(E)\nonumber\\ 
+ P_{12}^{++}(E)
&+&P_{12}^{--}(E)
\left.\rule{0cm}{0.5cm}\right]\,,
\label{eq3}
\end{eqnarray}
where all probabilities are evaluated at the same energy $E=eV/2$. Equation (\ref{eq3}) 
is easily derived from Eq.\ (\ref{eq3n}) using particle-hole symmetry 
$P_{ij}^{\alpha\beta}(E)=P_{ij}^{\bar\alpha\bar\beta}(-E)$ and 
flux conservation $N_i^\alpha(E)=\sum_{j\beta}P_{ij}^{\alpha\beta}(E)$.

In the Majorana scenario \cite{He294}, a $0.5e^2/h$ conductance 
results from all transmission probabilities in Eq.\ (\ref{eq3}) taking a value $\approx0.25$.
This is understood as an incident Fermionic quasiparticle  that splits and reflects in equal amounts 
as an electron and as a hole due to the peculiarity of the Majorana mode.
In presence of the magnetic orbital effect, a $0.5e^2/h$ conductance also results 
when one Fermion is fully transmitted, say $P_{12}^{++}=1$, while all the rest 
$P_{ij}^{\alpha\beta}$ vanish. This single-charged-mode scenario is 
made possible by the breaking of electron-hole degeneracy, which results in 
different transmissions for conjugate quasiparticles at the same energy
$P_{ij}^{\alpha\beta}(E)\ne P_{ij}^{\bar\alpha\bar\beta}(E)$. 
Nevertheless, we stress again that electron-hole symmetry is still
preserved by the orbital effect with the relations
$P_{ij}^{\alpha\beta}(E)=P_{ij}^{\bar\alpha\bar\beta}(-E)$.

The specific calculations of NSN topological junctions discussed below are performed with
the complex-$k$ method used previously by us in Refs.\ 
\citenum{Serra13,Osca15,Osca2017,Osca2017b} to solve the Bogoliubov-deGennes equation. More specifically, 
the reader is addressed to Ref.\ \citenum{Osca18b} for the details in the present context
of hybrid QAHI strips. Our approach allows to take fully into account the finite size influence
connected with specific values of $L_x$ and $L_y$ (Fig.\ \ref{Fig1}a).

\section{Results}
Figure \ref{Fig1} shows how the above two scenarios of $0.5e^2/h$ conductance can be easily distinguished by looking at the bias dependence $G(E)$ of an NSN double junction.  
 In presence of a Majorana zero mode (Fig.\ \ref{Fig1}b,c) the small bias region is characterized by 
 $G(E)\approx 0.5 e^2/h$, with small oscillations around this value due to the 
 interference effect  discussed in Ref.\ \citenum{Osca18b}.
On the contrary,  the charged-Fermion modes (Fig.\ \ref{Fig1}d,e) yield a 
$0.5e^2/h$ conductance plateau at finite bias that evolves, in the small bias limit, to an
$e^2/h$ conductance plateau. The reason of this difference at small bias is clear from Fig.\ \ref{Fig1}d: as $E$ decreases from positive values to zero the inverted band is activated, 
allowing the transport by the two Fermion modes of opposite charge
($P_{12}^{++}\approx P_{12}^{--}\approx 1$)
and a value $G\approx e^2/h$ from Eq.\ (\ref{eq3}).

For intermediate energies the halved conductance is found in both scenarios of Fig.\ \ref{Fig1}
and, therefore, it may become complicate to ascertain the $0.5e^2/h$ origin looking only at a single bias, or at a 
reduced window of biases. In this respect, Fig.\ \ref{Fig1} suggests that the conductance oscillations around
$0.5e^2/h$ are a differentiating mechanism in strips of $L_y$ smaller than a few microns.
Indeed, the conductance oscillations are due to the interference of the back-scattered chiral Majoranas\cite{Osca18b}, a mechanism that is absent in the transport by charged Fermion modes. 

\begin{figure}[t]
\begin{center}
\includegraphics[width=0.47\textwidth,trim=1.cm 10.cm 3.cm 3.cm,clip]{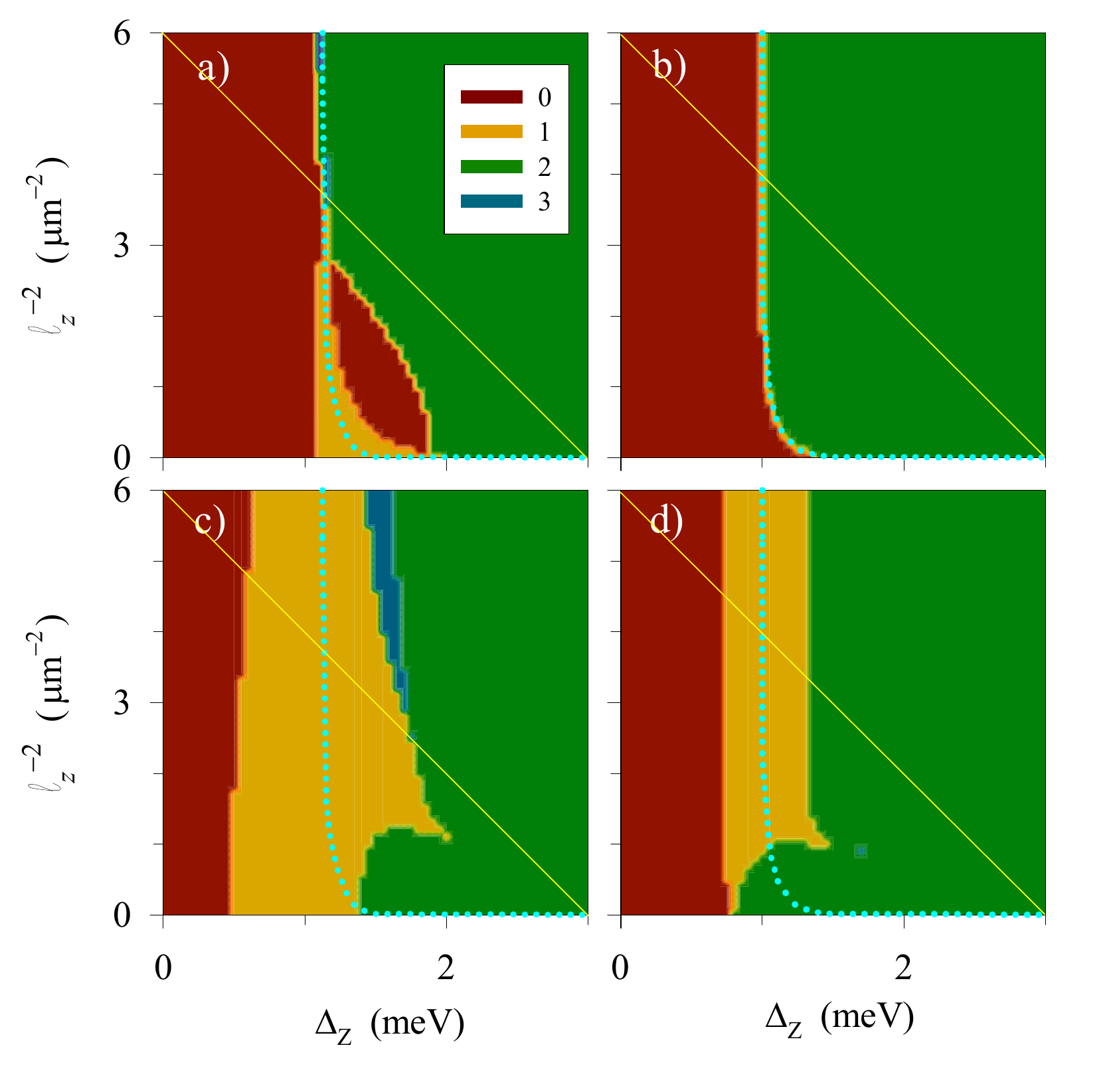}
\vspace{-3mm}
\end{center}
\caption{
Colorscale plots of the number of propagating modes ${\cal N}$ at a fixed energy $E$. 
We used $E=0.01$ meV (a,b) and $E=0.3$ meV (c,d). 
Left panels (a,c) are for the superconducting strip with $\Delta_1=1$ meV and 
$\Delta_2=0.1$ meV; right ones (b,d) are for $\Delta_{1,2}=0$.
Other parameters as in Fig.\ \ref{Fig1}.  
Dotted lines show the position of the $k=0$ gap closings while thin straight lines
represent a typical evolution in a  sweep of magnetic fields. 
}
\vspace{-5mm}
\label{Fig2}
\end{figure}

The phase diagrams with the number of propagating edge modes  
at low energies (${\cal N}$)
are shown in Fig.\ \ref{Fig2}
for selected values of the energy. In the limit of vanishing energy 
${\cal N}$ is the characteristic Chern number 
of each topological phase.    
At nonzero (sizeable) energies the phases
with ${\cal N}=1$ become prominent, extending up to relatively large 
magnetic fields
(Fig.\ \ref{Fig2}c,d). For $E\to 0$, however, the ${\cal N}=1$ phase is severely 
quenched, surviving only in a small region at low fields and with
nonvanishing superconductivity (Fig.\ \ref{Fig2}a).

It is remarkable that the ${\cal N}=1$ phase 
of an edge charged Fermion 
can be seen
even with vanishing superconductivity  (Fig.\ \ref{Fig2}d), although it requires 
the presence of magnetic field (${\ell_z^{-2}}>0$)
and nonvanishing energy (bias). This is showing that the 
essential requirements for this phase are particle-hole symmetry and breaking
of particle-hole degeneracy as discussed above, rather than a strong  superconducting
gap $\Delta_{1,2}$. The superconducting gap is nevertheless essential for the single 
Majorana phase, restricted to $\ell_z^{-2}\approx 0$.
Dotted lines in Fig.\ \ref{Fig2} mark the position of the $k=0$ energy gap closing.
Ideally, this line separates the energy-band behavior from a 
gapped band structure to a double cross at non zero $k$'s, i.e., 
a transition ${\cal N}=0\to{\cal N}=2$.
The separation in $k$ values between the two crosses 
increases with 
the magnetic field, $\Delta k\propto L_y\ell_z^{-2}$ and, only for 
$\ell_z^{-2}\approx 0$ a single cross at the origin is effectively present.
This agrees with Fig.\ \ref{Fig2}b and with Fig.\ \ref{Fig2}a for $\ell_z^{-2}> 3\, \mu{\rm m}^{-2}$.
With nonzero $E$'s and $\ell_z^{-2}$'s the intermediate phase ${\cal N}=1$ may still emerge 
either because $E$ exceeds the $k=0$ maximum (Fig.\ \ref{Fig1}d and Fig.\ \ref{Fig2}c,d), or
because a small gap is still present at finite $k$ before the double cross is fully formed 
(Fig.\ \ref{Fig2}a). The latter, however, is a marginal finite-$L_y$ 
effect yielding the severely 
quenched ${\cal N}=1$ phase in Fig.\ \ref{Fig2}a.

\begin{figure}[t]
\begin{center}
\includegraphics[width=0.4\textwidth,trim=1.5cm 12.cm 4.cm 5.5cm,clip]{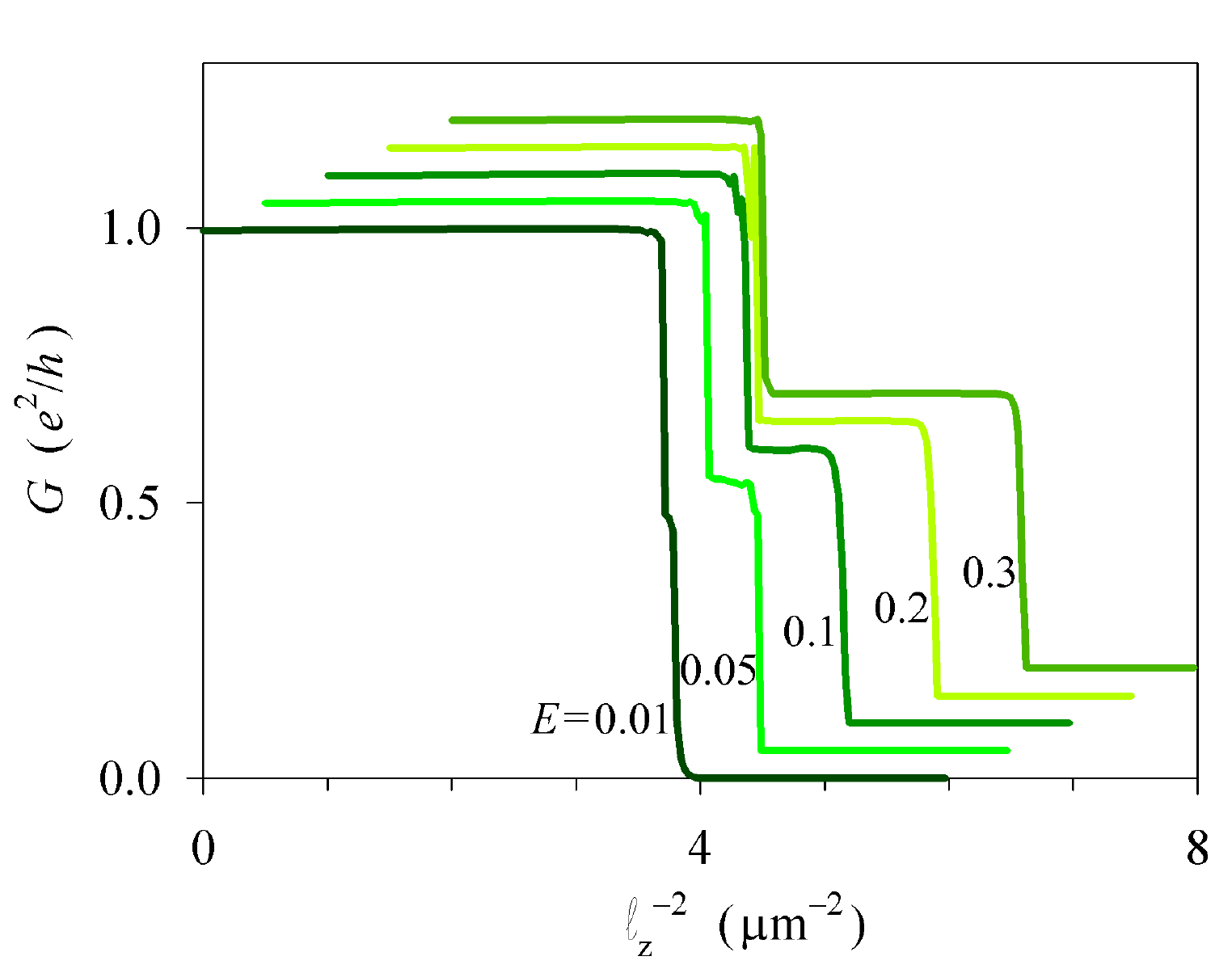}
\vspace{-3mm}
\end{center}
\caption{
Conductance traces for a magnetic field sweep along the straight lines 
shown in Fig.\ \ref{Fig2}. The bias (energy) is kept fixed in each trace to the 
indicated value. For clarity, each trace has been shifted with respect to the preceding one, in 
increasing energy order, by an increment of 0.5 horizontally and 0.05 vertically.
The $0.5e^2/h$ plateau is greatly enhanced with increasing bias.
}
\vspace{-5mm}
\label{Fig3}
\end{figure}

We have explored above the model dependence on $\Delta_Z$ and $\ell_z^{-2}$ assumed to 
be independent parameters.
In experiment\cite{He294}, however, the material magnetization ($\Delta_Z$) is triggered 
by the magnetic field ($B_z$ or, equivalently, $\ell_z^{-2}$) in between saturation values. Noting the  symmetry properties
\begin{eqnarray}
G(E,\Delta_Z,B_z)
&=&
G(E,-\Delta_Z,B_z)\nonumber\\
&=&
G(E,\Delta_Z,-B_z)\nonumber\\
&=&
G(E,-\Delta_Z,-B_z)\; ,
\end{eqnarray}
it is always possible to represent a general trajectory in the $\Delta_Z-B_z$ plane by
its corresponding image in the first quadrant of the plane. In this respect, the thin diagonal 
lines of Fig.\ \ref{Fig2} are a possible linear dependence $\Delta_Z-\ell_z^{-2}$ modelling
a sweep of magnetic field from negative to positive values in a magnetization reversal.
The corresponding conductances can be seen in Fig.\ \ref{Fig3} for different values of the energy (bias).

The conductance traces of Fig.\ \ref{Fig3} show that the $0.5e^2/h$ plateau becomes conspicuous 
with increasing bias as a result of the magnetic orbital effect. In our model this bias dependence is to be expected whenever the conductance transition and its associated plateau occurs at magnetic fields
$\ell_z^{-2}\gtrsim 1 $ $\mu{\rm m}^{-2}$, or equivalently $B_z\gtrsim 0.7$ mT. As argued above, 
in this scenario the $0.5e^2/h$ plateau reflects the breaking of electron-hole degeneracy by the
magnetic orbital motion with a given novanishing energy.
It is not to be excluded, however, that the $0.5e^2/h$ plateau due to a genuine Majorana mode can be observed, although our above analysis indicates that this would require rather small magnetic 
fields such that the field sweep in Fig.\ \ref{Fig2} moves towards the horizontal axis.
We also stress that in the Majorana scenario the $0.5$ plateau should still survive in the limit of vanishing bias $E\to 0$; therefore, observing a clear bias dependence as in Fig.\ 3 would discern 
the validity of the charged Fermion scenario suggested here.

\section{Conclusions}

A $0.5e^2/h$ conductance plateau in ideal quantum-anomalous-Hall hybrid strips can be attributed to
a chiral Majorana mode at very small magnetic fields, or to a charged Fermionic 
mode at sizeable fields ($B_z>1\, {\rm mT}$).
For the latter, the magnetic orbital motion has to break the electron-hole degeneracy  at a given finite energy,
while still keeping electron-hole symmetry in the transmissions for opposite-sign energies. Both  situations can be 
realized in a two terminal setup with a grounded superconductor region and a symmetrical
bias drop in the two terminals. The vanishing-bias dependence of the $0.5$ plateau, disappearing 
for the charged Fermion  and surviving for the chiral Majorana, has been 
suggested as a means to discern the two scenarios of a halved conductance.
In addition, while the 0.5 plateau due to a Majorana is predicted to display energy oscillations 
when the transverse width is small $L_y\lesssim 3$ $\mu$m, \cite{Osca18b} the plateau due to a
charged Fermion is predicted to be much more $L_y$-independent and flat.
The latter is characterized by an almost perfect transmission with magnetic motion along the edge,  largely insensitive to $L_y$.

\begin{acknowledgement}
This work was funded by MINEICO (Spain), grant MAT2017-82639.
\end{acknowledgement}

\bibliography{naleorb}

\end{document}